\documentclass{article}
\usepackage[a4paper, total={7in, 10in}]{geometry}

\usepackage[utf8]{inputenc} 
\usepackage[T1]{fontenc}    
\usepackage{hyperref}       
\usepackage{url}            
\usepackage{booktabs}       
\usepackage{amsfonts}       
\usepackage{nicefrac}       
\usepackage{microtype}      
\usepackage{lipsum}
\usepackage{amsmath,mathtools,bbm,multirow,url,hyperref,enumerate}
\usepackage{color}
\usepackage{float}
\usepackage{hyperref}
\usepackage{makecell}
\usepackage{algorithm,algorithmic}
\usepackage{multicol}
\usepackage{natbib}
\usepackage{hyperref}
\usepackage{soul}
\hypersetup{colorlinks,citecolor=red}
\newcommand{\com}[1]{\textcolor{black}{#1}}
\usepackage{setspace} 

\title{A Bayesian design for dual-agent dose optimization with targeted therapies}

\author{
  Jos\'e L. Jim\'enez \\
  Novartis Pharma A.G., Basel, Switzerland \\
  \texttt{jose\_luis.jimenez@novartis.com} \\
  Mourad Tighiouart \\
  Cedars-Sinai Medical Center, Los Angeles, USA \\
  \texttt{Mourad.Tighiouart@cshs.org}
}

\begin{document}
\maketitle

\begin{abstract}
In this article, we propose a phase I-II design in two stages for the combination of molecularly targeted therapies. The design is motivated by a published case study that combines a MEK and a PIK3CA inhibitors; a setting in which higher dose levels do not necessarily translate into higher efficacy responses. The goal is therefore to identify dose combination(s) with a prespecified desirable risk-benefit trade-off. We propose a flexible cubic spline to model the marginal distribution of the efficacy response. In stage I, patients are allocated following the escalation with overdose control (EWOC) principle whereas, in stage II, we adaptively randomize patients to the available experimental dose combinations based on the continuously updated model parameters. A simulation study is presented to assess the design's performance under different scenarios, as well as to evaluate its sensitivity to the sample size and to model misspecification. Compared to a recently published dose finding algorithm for biologic drugs, our design is safer and more efficient at identifying optimal dose combinations.
\end{abstract}

\section{Introduction}
\label{sc_introduction}

Integrated phase I-II clinical trial designs allow accelerated drug development since they assess, within a single protocol, the safety and efficacy of a compound or a combination of drugs. With the use of traditional chemotherapy compounds, a dose limiting toxicity (DLT) is generally ascertained after one cycle of therapy for the purpose of estimating the maximum tolerated dose (MTD), and it is generally assumed that both the dose-toxicity and dose-efficacy relationships are monotonically increasing functions (see e.g., \cite{le2009dose}). This implies that the optimal dose, i.e., the dose with most desirable benefit-risk trade-off, must be at the MTD. However, as pointed out by \cite{hoff2007targeted,li2017toxicity,lin2020adaptive}, with other types of compounds such us molecularly targeted therapies or immunotherapies, the monotonicity assumption of of the dose-efficacy response may not hold given that efficacy may plateau or decrease at high dose levels, which implies that the optimal dose may not be located at the MTD.

Phase I-II designs that do not assume monotonicity of the dose-efficacy relationship have been studied extensively in the last two decades. For instance, \cite{thall2004dose}, proposed a Bayesian phase I-II design based on toxicity-efficacy probability trade-offs. \cite{nebiyou2005bayesian} proposed a design in which binary toxicity outcomes and a continuous biomarker expression outcomes are jointly modeled. \cite{zhang2006adaptive} employed a flexible continuation-ratio model to account for the potentially monotonically increasing / non-increasing / decreasing dose-efficacy profiles. \cite{houede2010utility} proposed a design for drug combinations with ordinal toxicity and efficacy outcomes in which the optimal dose combination was found through a utility function. \cite{yuan2011bayesian} proposed a phase I-II design for late-onset efficacy with drug combinations that incorporates adaptive randomization of patients in stage II with the intention of allocating more patients to more efficacious dose combination levels. \cite{cai2014bayesian} proposed a phase I-II design in two stages in which the optimal dose combination is estimated by encouraging the exploration of untried dose combinations to avoid the problem of recommending suboptimal doses. \cite{guo2017bayesian} proposed a phase I-II design for molecularly targeted agents that considers different biomarker subgroups. \com{Recently, \cite{lu2024comb} published the ``Comb-BOIN12'', the drug combination version of the phase I-II BOIN design ``BOIN12'' (\cite{lin2020boin12}). This model-assisted design is particularly interesting since it is accurate while simplifying the decision-making during interim periods.} \cite{lyu2019aaa} proposed a two-stage phase I-II design in which the optimal dose combination is obtained my maximizing a utility function. One characteristic of this design is that it allows to do adaptive dose insertion if the current estimate of the optimal dose combination is far from all the pre-defined (discrete) dose combination levels available in the trial. We find this feature particularly appealing in the context of molecularly targeted therapies because it allows to adapt the initial grid of discrete dose combination levels, if necessary. Without it, we may incur in a substantial loss of information, especially when the knowledge about the dose-toxicity and the dose-efficacy surfaces is limited (\cite{diniz2019comparison}). However, adaptive dose insertion may be challenging in practice since new drug formulation may be frequently requested with short notice. We note that adaptive dose insertion is conceptually similar to having continuous dose levels administered intravenously, which has been extensively studied in both phase I and phase I-II designs by \cite{tighiouart2014,tighiouart2017bayesian,diniz2017,diniz2018,jimenez2019,tighiouart2019two,jimenez2020bayesian,jimenez2021combining,jimenez2021bayesian}, in the setting of cytotoxic agents.

Our research is motivated by a published phase I-II clinical trial design that combines a MEK and a PIK3CA inhibitors (\cite{lyu2019aaa}), considering four discrete dose levels for each compound, and enrolling a total of 96 late-stage cancer patients. The primary endpoint of the study was to improve the efficacy rate from 5\% to 30\% taking into consideration that a dose combination with 20\% efficacy rate or higher is consider beneficial as long as the dose is well tolerated. 

In this article, we propose a two-dimensional flexible cubic spline function to model the marginal distribution of the efficacy response in settings combining either two molecularly targeted agents or a cytotoxic with a molecularly targeted agent. In stage I, the estimated MTD is calculated using the escalation with overdose control (EWOC) principle. In stage II, instead of allocating patients directly to the standardized dose combination with the currently highest utility estimate, we follow the adaptive randomization principle, which prevents the design to become stuck at local optima (\cite{yuan2016bayesian}). The most desirable benefit-risk trade-off (i.e., the optimal dose combination) is calculated by maximizing a utility function.

The remainder of this article is organized as follows. In section \ref{sc_method} we introduce the dose-toxicity and dose-efficacy models together with the utility function and the stage I and stage II dose finding algorithms. In section \ref{sc_simulation_study}, we present an extensive simulation study to evaluate the operating characteristics of the approach as well as a comparison with state-of-the-art methodology in the context of this type of clinical trial. We conclude the article in section \ref{sc_discussion} with a discussion and some concluding remarks.

\section{Method}
\label{sc_method}
\subsection{Probability models}

Consider a phase I-II design combining $Q$ pre-specified doses $x_1 < \dots < x_Q$ of compound $X$ with $\mathcal{J}$ pre-specified doses $y_1 < \dots < y_{\mathcal{J}}$ of compound $Y$. The dose levels of each compound are standardized to fall within the interval [0,1] using the transformations $f_x(x) = (x - x_1) / (x_Q - x_1)$ and $f_y(y)  = (y - y_1) / (y_{\mathcal{J}} - y_1)$, with $x = x_1, \dots, x_Q$ and $y = y_1, \dots, y_{\mathcal{J}}$, resulting into $x \in [0,1]$ and $y \in [0,1]$, respectively. Let $Z_T = \{0,1\}$ be the binary indicator of DLT where $Z_T = 1$ represents the presence of a DLT after a predefined number of treatment cycles, and $Z_T = 0$ otherwise. Let $Z_E = \{0,1\}$ be the binary indicator of treatment response where $Z_E = 1$ represents a positive response after a predefined number of treatment cycles, and $Z_E = 0$ otherwise.

The optimal dose combination as well as all the adaptive features of the design are based on the joint probability $P(\textbf{Z} | x, y, \boldsymbol{\Psi})$, where $\textbf{Z} = \{Z_T, Z_E\}$ is a vector containing the toxicity and efficacy binary outcomes, and $\boldsymbol{\Psi} = \{\boldsymbol{\Psi}_T, \boldsymbol{\Psi}_E\}$ is a vector containing the parameters of the marginal toxicity and efficacy models, respectively. For notational simplicity, we suppress the arguments $\boldsymbol{\Psi}_T$ and $\boldsymbol{\Psi}_E$ when it will not cause confusion. Following the work of \cite{ivanova2009adaptive,cai2014bayesian,lyu2019aaa,jimenez2020bayesian,jimenez2021combining} we model the marginal toxicity and efficacy models independently (i.e., we assume that toxicity and efficacy are independent). The marginal probability of toxicity $P(Z_T = 1 | x, y)$ is modeled using the linear logistic regression model
\begin{equation}
\label{eq_pdlt}
     P(Z_T = 1 | x,y) = F(\alpha_0 + \alpha_1 x + \alpha_2 y + \eta x y),
\end{equation}where $F(.)$ is the cumulative distribution function of the logistic distribution (i.e., $F(u) = 1 / (1 + e^{-u})$). Following \cite{tighiouart2017bayesian}, we reparameterize equation \eqref{eq_pdlt} in terms of parameters that clinicians can easily interpret. Let $\rho_{uv}$ denote the probability of DLT when the levels of agents $X = u$ and $Y = v$, with $u = \{0, 1\}$, and $v = \{0, 1\}$, so that $\alpha_0 = F^{-1}(\rho_{00})$, $\alpha_1 = (F^{-1}(\rho_{10}) - F^{-1}(\rho_{00}))$, and $\alpha_2 = (F^{-1}(\rho_{01}) - F^{-1}(\rho_{00}))$. \com{In other words, $\rho_{00}$ represents the probability of DLT with the lowest dose levels of both X and Y, $\rho_{01}$ represents the probability of DLT with the lowest dose level of X and the highest dose level of Y, and $\rho_{10}$ represents the probability of DLT with the highest dose level of X and the lowest dose level of Y}.

The marginal probability of efficacy $P(Z_E = 1 | x, y)$ is modeled using the cubic spline
\begin{equation}
\label{eq_peff}
    P(Z_E = 1 | x, y) = F(\beta_0 + \beta_1 x + \beta_2 x^2 + \sum_{i=3}^{5} \beta_i (x - \kappa_{i-2})_{+}^{3} + \beta_6 y + \beta_7 y^2 + \sum_{j=8}^{10} \beta_j (y - \kappa_{j-4})_{+}^{3} + \beta_{11}xy),
\end{equation}where $\kappa_1 = \kappa_4 = 0$.

To shorten the notation, let $P(Z_T = 1 | x,y) = \pi_T(x,y)$ and $P(Z_E = 1 | x, y) = \pi_E(x,y)$.

\subsection{Prior distributions}

The prior distribution of the parameters in $\pi_T(x,y)$ and $\pi_E(x,y)$ are usually elicited after consultation with the clinicians based on previous single agent and/or drug combination studies. 

In this article, we do not have any elicited prior distribution and therefore we use the following vague distributions: $\rho_{01} \sim \mbox{beta}(1,1)$, $\rho_{10} \sim \mbox{beta}(1,1)$, and conditional on $(\rho_{01}, \rho_{10})$, we assume that $\rho_{00} / \min (\rho_{01}, \rho_{10}) \sim \mbox{beta}(1,1)$. The interaction parameter $\eta$ represents the  synergism of the combination which means that it has to be positive. We assign $\eta$ a vague gamma prior, for example $\eta \sim \mbox{gamma(0.1, 0.1)}$.

In $\pi_E(x,y)$, we assume vague normal distributions for all the $\beta$ parameters so that $\beta_r \sim N(0,10^2)$, $r = 1, \dots, 11$. The justification for allowing all the $\beta$ parameters to be either positive or negative is that we expect the dose-efficacy surface to be non-linear, which implies that efficacy may decrease at higher dose combination levels. We assign $\kappa_2,\kappa_3,\kappa_5,\kappa_6$ uniform distributions with the logical restriction that $\kappa_2 < \kappa_3$ and $\kappa_5 < \kappa_6$. Thus, $(\kappa_2,\kappa_3) \sim \mbox{Uniform}((\kappa_2,\kappa_3): 0 \leq \kappa_2 < \kappa_3 \leq 1)$ and $(\kappa_5,\kappa_6) \sim \mbox{Uniform}((\kappa_5,\kappa_6): 0 \leq \kappa_5 < \kappa_6 \leq 1)$. Conventionally, vague priors have enormous variances, which work well with medium to large sample sizes but may lead to numerical instability with smaller sample sizes. Ideally, the prior distributions should be vague enough to cover all the plausible values of the parameters, but not too vague that causes stability issues.

\subsection{Likelihood and posterior distributions}

Let $N$ denote the maximum sample of the trial and $D_n = { (Z_{T_i}, Z_{E_i}, x_i, y_i), i = 1, \dots, n }$ be the data collected (i.e., toxicity outcomes, efficacy outcomes and dose combinations) after enrolling $n$ patients.

The posterior distribution of the dose-toxicity model parameters is $p({\boldsymbol \Psi}_T | D_n) \propto p({\boldsymbol \Psi}_T) \times \mathcal{L}(D_n | {\boldsymbol \Psi}_T)$, and the posterior distribution of the dose-efficacy model parameters is $p({\boldsymbol \Psi}_E | D_n) \propto p({\boldsymbol \Psi}_E | D_n) \times \mathcal{L}(D_n | {\boldsymbol \Psi}_E)$, where ${\boldsymbol \Psi}_T = (\rho_{00}, \rho_{01}, \rho_{10}, \eta)$, ${\boldsymbol \Psi}_E = (\beta_0, \dots, \beta_{11}, \kappa_1, \dots, \kappa_6)$, $\mathcal{L}(D_n | {\boldsymbol \Psi}_T) = \prod_{i=1}^{m} \pi_T(x_i,y_i)^{Z_i} \times (1 -\pi_T(x_i,y_i))^{1-Z_i}$ and $ \mathcal{L}(D_n | {\boldsymbol \Psi}_E) = \prod_{i=1}^{n} (\pi_E(x_i,y_i))^{E_i} \times (1 - \pi_E(x_i,y_i))^{1-E_i}$. Bayesian computation is done using \texttt{JAGS} (\cite{plummer2003jags}) and \texttt{R} (\cite{r2019}).

\subsection{Optimal dose combination}
\label{sc_bodc}

As mentioned in section \ref{sc_introduction}, with molecularly targeted therapies it is not guaranteed that the optimal dose combination will be at the MTD, which means we will need to find the most desirable trade-off between risk and benefit. Ideally, we would like a dose combination with low toxicity and high efficacy (i.e., low risk and high benefit). 
Utility functions $U(.)$ are convenient tools that allow to formally assess the benefit-risk trade-off between undesirable and desirable clinical outcomes. They have received considerable attention over the last years, specially with the development of targeted therapies and immuno-therapies (see e.g., \cite{houede2010utility, thall2013using, thall2014optimizing, guo2015bayesian, murray2017robust,liu2018bayesian,lyu2019aaa}), and their definition can be more or less complicated, depending on the setting and the number of outcomes that we may involve in the trade-off. For example, \cite{liu2018bayesian} considers three outcomes (toxicity, efficacy and immune response). Once the utility function is defined, the optimal dose (combination) will correspond to the dose (combination) that maximizes the utility function based on the current parameters estimates. In section \ref{sc_simulation_study}, we provide the definition of the utility function used for the simulation study, which in generic terms, we refer to as $U(x,y)$. However, each trial will have a different utility function based on the clinicians' criteria about benefit-risk trade-off desirability.




\subsection{Dose-optimization algorithm}

We define the MTD as any dose combination with
\begin{equation}
\label{eq_mtd}
F(\alpha_0 + \alpha_1 x + \alpha_2 y + \eta x y) = \theta_T,
\end{equation}where $\theta_T$ represents the highest probability of toxicity we are willing to accept. With discrete dose levels, it is possible that none of the experimental dose levels have an exact probability of toxicity equal to $\theta_T$ and therefore the MTD could be an empty set. 

In this article, stage I follows the escalation with overdose control (EWOC) principle (\cite{babb1998cancer, tighiouart2005flexible, tighiouart2010dose, tighiouart2017bayesian, tighiouart2012number, shi2013escalation}), where the posterior probability of overdosing the next cohort of patients is bounded by a feasibility bound $\alpha$. 
In a cohort with two patients, the first one would receive a new dose of compound $X$ given that the dose $y$ of compound $Y$ that was previously assigned. The other patient would receive a new dose of compound $Y$ given that dose $x$ of compound $X$ was previously assigned. The feasibility bound $\alpha$ increases from 0.25 up to 0.5 in increments of 0.05. \com{For a detailed explanation of how the feasibility bound with $\alpha$ increments works, we encourage the reader to read the article (\cite{wheeler2017toxicity})}.

In other words and based on \eqref{eq_mtd}, if $x$ is fixed, the posterior distribution of the MTD takes the form $[\tilde{y}_{\mbox{\tiny MTD}}~|~\tilde{{\boldsymbol \Psi}}_T,x] = (F^{-1}(\theta_T) - \tilde{\alpha}_0 - \tilde{\alpha}_1 \times x) / (\tilde{\alpha}_2 + \tilde{\alpha}_3 \times x)$ whereas if $y$ is fixed, it takes the form $[\tilde{x}_{\mbox{\tiny MTD}}~|~\tilde{{\boldsymbol \Psi}}_T,y] =  (F^{-1}(\theta_T) - \tilde{\alpha}_0 - \tilde{\alpha}_2 \times y) / (\tilde{\alpha}_1 + \tilde{\alpha}_3 \times y)$, with $\tilde{{\boldsymbol \Psi}}_T = \{\tilde{\alpha}_0 = F^{-1}(\tilde{\rho}_{00}), \tilde{\alpha}_1 = (F^{-1}(\tilde{\rho}_{10}) - F^{-1}(\tilde{\rho}_{00})), \tilde{\alpha}_2 = (F^{-1}(\tilde{\rho}_{01}) - F^{-1}(\tilde{\rho}_{00})), \tilde{\eta}_3\}$ representing the current posterior distribution of the model parameters in \eqref{eq_pdlt}.




Stage I will enroll a total of $N_1 = C_1 \times m_1$ patients, where $C_1$ represents the total number of cohorts in phase I with equal number of patients $m_1$. \com{Even though $m_1$ could be any number considered reasonable, we consider throughout the article $m_1 = 2$ and therefore the description of the algorithm is done assuming such cohort size}. The stage I algorithm proceeds as follows:

\begin{enumerate}

\item The first cohort ($c_1 = 1$) of $m_1$ patients starts at the dose combination $(x = 0, y = 0)$. 

\item For cohorts $c_1 > 1$, if $c_1$ is an even number, 
\begin{itemize}
\item[i)] Patient $2c_{1} - 1$ receives the dose combination $(x_{2c_{1}-1}, y_{2c_{1}-1} = y_{2c_{1}-3})$ where $x_{2c_{1}-1}$ represents the discrete dose level in $\mathcal{Q}$ that is closest, in terms of euclidean distance, to the $\alpha$-th percentile of $[\tilde{x}_{\mbox{\tiny MTD}}~|~\tilde{{\boldsymbol \Psi}}_T,y = y_{2c_{1}-3}]$.

\item[ii)] Patient $2c_{1}$ receives the dose combination $(x_{2c_{1}} = x_{2c_{1}-2}, y_{2c_{1}})$ where $y_{2c_{1}}$ represents the discrete dose level in $\mathcal{J}$ that is closest, in terms of euclidean distance, to the $\alpha$-th percentile of $[\tilde{y}_{\mbox{\tiny MTD}}~|~\tilde{{\boldsymbol \Psi}}_T,x = x_{2c_{1}-2}]$.
\end{itemize}

In contrast, if $c_1$ is an odd number, 
\begin{itemize}
    \item[i)] Patient $2c_{1} - 1$ receives the dose combination $(x_{2c_{1} - 1} = x_{2c_{1} - 3}, y_{2c_{1} - 1})$ where $y_{2c_{1} - 1}$ represents the discrete dose level in $\mathcal{J}$ that is closest to the $\alpha$-th percentile of $[\tilde{y}_{\mbox{\tiny MTD}}~|~\tilde{{\boldsymbol \Psi}}_T,x = x_{2c_{1}-3}]$.
    
    \item[ii)] Patient $2c_{1}$ receives the dose combination $(x_{2c_{1}}, y_{2c_{1}} = y_{2c_{1} - 2})$ where $x_{2c_{1}}$ represents the discrete dose level in $\mathcal{Q}$ that is closest to the $\alpha$-th percentile of $[\tilde{x}_{\mbox{\tiny MTD}}~|~\tilde{{\boldsymbol \Psi}}_T,y = y_{2c_{1}-2}]$.
    
\end{itemize}

\item Keep enrolling cohorts of size $m_1$ until $c_1 = C_1$.
\end{enumerate}


\com{We refer the reader to Figure S1 in the supplementary material of \cite{jimenez2020bayesian} for a graphical representation of the dose-escalation algorithm in stage I.} 

Stage II will enroll a total of $N_2 = n_2 + C_2 \times m_2$ patients, where $m_2$ represents the cohort size and $n_2$ represents the initial fixed cohort of patients in which the probability of  allocating a patient to a dose combination is the same along the space of safe dose combinations. \com{The rational for the initial cohort of patients $n_2$ is to increase the amount of efficacy data available in doses considered safe after stage I.} Let $N = N_1 + N_2$ be the total number of patients that the entire study will enroll. The algorithm in stage II proceeds as follows:

\begin{enumerate}

    \item An initial cohort ($c_2 = 1$) of size $n_2$ is distributed \com{homogeneously (i.e., with equal probability)} among the dose combinations with $\widehat{\pi}_T(x,y) \leq \theta_T$ given $D_{N_1}$, so that as many dose levels as possible have at least one patient allocated to them. 
    
    \item For cohorts $c_2 > 1$ of size $m_2$, we sequentially allocate patients using adaptive randomization in which the probability of being allocated to a dose combination is proportional to the current utility estimate (i.e., $\pi_{\mbox{\tiny AR}}(x,y) = \widehat{U}(x,y) / \sum \widehat{U}(x,y)$, where $\pi_{\mbox{\tiny AR}}(x,y)$ represents the probability of being allocated to dose combination $(x,y)$).
    
    
    \item Keep enrolling cohorts of size $m_2$ until $c_2 = C_2$.
    
\end{enumerate}

So far, the design has only used the initial set of discrete dose combinations. One important question is whether we should restrict the optimal dose combination recommendation to the initial grid of discrete dose combination, or whether we should allow the design to recommend any (continuous) dose combination within the standardized space of dose combinations $[0,1] \times [0,1]$. We believe that the design is flexible enough to identify the region (or regions) in which the true utility is higher, even in settings with complex dose-efficacy surfaces. Thus, having the option to recommend a continuous dose combination would improve the chances of identifying the true optimal dose combination in case it would be distant from all of the existing discrete dose combination levels. Therefore, at the end of stage II, the optimal dose combination is calculated as  
\begin{equation}
    (x^{\mbox{\tiny OPT}},y^{\mbox{\tiny OPT}}) = \underset{(x,y) \in [0,1]\times [0,1]}{\mbox{arg max}} \left \{ \widehat{U}(x,y) \right \}.
\end{equation}

\com{It is important to acknowledge that even though it is desirable to have the flexibility of recommending a dose combination outside the initial dose combination grid, this can also imply that we would be declaring as optimal a dose combination at which no patients have been treated. In this case, we recommend that clinicians treat a small number of patients at this recommended dose sequentially using a Bayesian monitoring stopping rule for safety.}

The proposed design contains two stopping rules for safety, one for stage I and a different one for stage II. During stage I, we would stop the trial if
\begin{equation}
   P(\pi_T(x = 0,y = 0) > \theta_T + 0.1| D_n) > \delta_{\theta_1},
\end{equation}where $\delta_{\theta_1} = 0.5$. In contrast, during stage II we would stop the trial if
\begin{equation}
   P(\Theta > \theta_T + 0.1|D_n) > \delta_{\theta_2},
\end{equation} where $\Theta$ represents the rate of DLTs for both stages of the design regardless of dose and $\delta_{\theta_2} = 0.7$ represents the confidence level (i.e., 70\%) that a prospective trial results in an excessive DLT rate. A non-informative Jeffreys prior Beta$(0.5,0.5)$ is placed on the parameter $\Theta$.

Stopping rules for futility and efficacy are not considered in this trial given the potential complexity of the dose-efficacy surface and relatively small sample size. However, if deemed necessary, they could be incorporated into the design.


\section{Simulation Studies}
\label{sc_simulation_study}

In this section, we describe the performance of our approach in identifying the optimal dose combination, compare the safety of the trial and average utility of the recommended dose combinations with the AAA design, and study robustness to varying sample size.

\subsection{Model and design performance}
\label{sc_model_performance}

We assess the performance of our design using simulation studies. Considering the setting of the motivating trial, we assume that both compounds $X$ and $Y$ have four standardized dose levels within the interval $[0,1]$. In stage I we select $m_1 = 2$ and $C_1 = 15$, yielding a total of $N_1 = 30$ patients. In stage II we select $n_2 = 12$, $m_2 = 6$ and $C_2 = 9$, yielding a total of $N_2 = 66$ patients. In order to compare the performance of our approach with the triple adaptive (AAA) Bayesian design of \cite{lyu2019aaa}, we let the toxicity upper bound $\theta_T = 0.3$, the efficacy lower bound $\theta_E = 0.2$ and the lowest acceptable utility defined below as $U_0 = 0.1$. The utility function used in \cite{lyu2019aaa} is defined as 
\begin{equation}
\label{eq_utility}
U(x,y) = \mathbf{1} \left ( \pi_T(x,y) \leq \theta_T \right ) \times \left ( 1 - \frac{(1 - \eta_0) \times \pi_T(x,y)}{\theta_T} \right ) \times [\eta_1 \exp(\eta_2 \times \pi_{E}(x,y)) + \eta_3],
\end{equation}where `$\mathbf{1}(.)$' denotes an indicator function and  $\eta_0 = 0.368$, $\eta_1 = 0.385$, $\eta_2 = 1.28$ and $\eta_3 = -0.385$ are parameters, elicited by the physicians, that establish the benefit-risk trade-offs (see \cite{lyu2019aaa}, sections 2.2 and 4.1). This utility function has the property that for fixed probability of efficacy $\pi_E(x,y)$ at dose combination $(x,y)$, it decreases linearly as a function of $\pi_T(x,y)$ provided that $\pi_T(x,y) \leq \theta_T$ and it increases exponentially as a function of $\pi_E(x,y)$ for fixed risk of DLT $\pi_T(x,y)$, provided that $\pi_T(x,y) \leq \theta_T$. \com{This utility function was selected based on an extensive discussion between the authors of \cite{lyu2019aaa} and the clinician and does depend on how much toxicity trade-off for efficacy the clinician is willing to accept. Other choices of futility functions may be explored and may lead to different operating characteristics.}

The design is evaluated through the following operating characteristics: i) average DLT rate and percentage of trials with DLT rate above $\theta_T$ and $\theta_T + 0.1$, ii) probability of early stopping for safety, iii) average (true) utility of the recommended optimal dose combinations, iv) average (true) utility of the patients allocated with adaptive randomization, and v) the distribution of the recommended optimal dose combinations.

\begin{figure}
\caption{Contour plots showing the utility surface from scenarios 1-6. We also display the grid of discrete dose combination levels (i.e., black dots) with their corresponding (true) utility values whenever these (i.e., the utility values) are above zero.}
\centering
\vspace{0.25cm}
\includegraphics[scale=0.7]{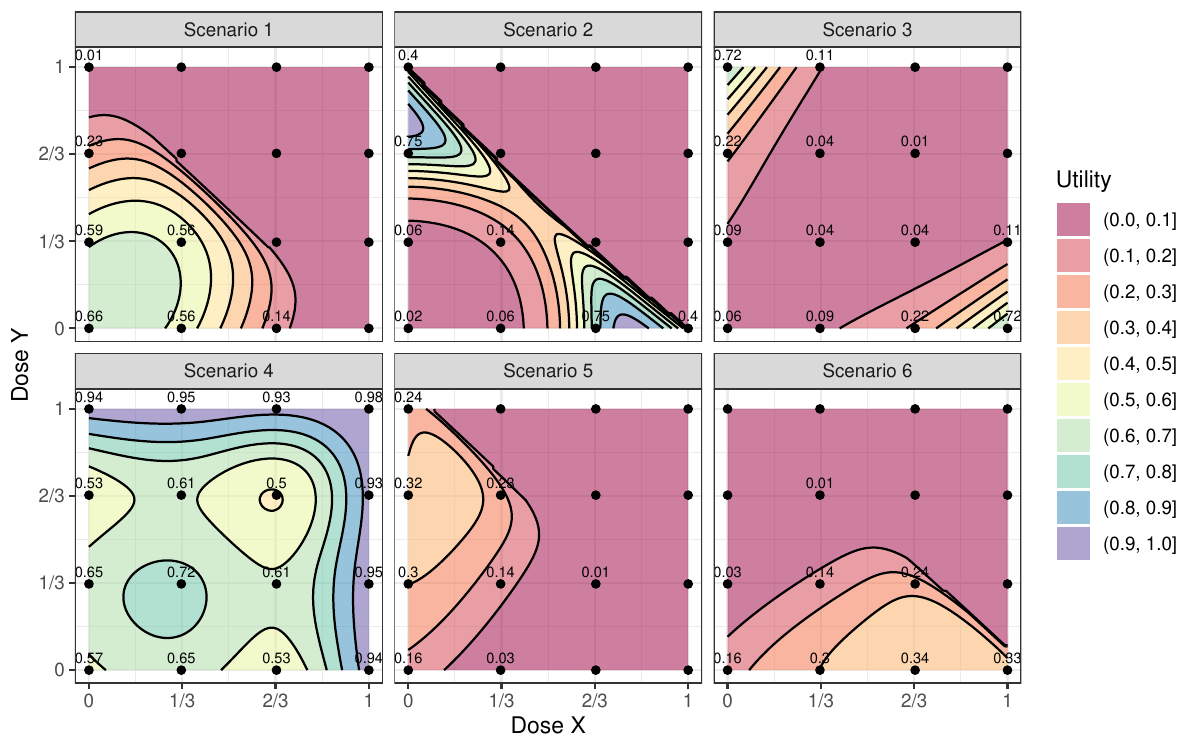}
\label{plot_simulation_scenarios}
\end{figure}

We specified 6 scenarios using the marginal models defined in equations \eqref{eq_pdlt} and \eqref{eq_peff} that vary with respect to the location of the true optimal dose combination (also referred as target dose combination) as well as in the complexity of surface of the utility function (i.e., uni-modal and bi-modal). Figure \ref{plot_simulation_scenarios} shows the (true) utility surfaces for these scenarios. The parameter values used to produce these scenarios are available in Table S1 of the supplementary material. Under each scenario, we simulated 2000 trials. To assess the performance of our design relative to the state-of-the-arm AAA methodology in \cite{lyu2019aaa}, we compare the average, median and the 2.5th and 97.5th percentiles of the distribution of the (true) utilities of the recommended optimal dose combinations of these two designs. In the remainder of the manuscript, we refer to the 2.5th and 97.5th percentiles of the (empirical) distribution of the (true) utilities of the recommended optimal dose combinations as a 95\% confidence interval, which should interpreted as a measure of precision or reliability. Also, the (true) utilities of the recommended optimal dose combination are referred to recommended (true) utilities.

Scenarios 1, 5 and 6 are designed to be uni-modal whereas scenarios 2, 3 and 4 are designed to be bi-modal and therefore more complex. The goal is to have scenarios with low, medium and high utility values for the target dose combination. Scenario 3 is expected to be particularly challenging. In this scenario, we placed very low utility values in the entire path that stage I is expected to follow given the very low toxicity probability that we established throughout the entire dose-toxicity surface. At the end of stage I, the number of positive efficacy responses is expected to be very low and stage II will start with very little information regarding the potential location of the target dose combination.

In these scenarios, the average DLT rate ranged between 7-17\%, the proportion of trials with DLT rates above $\theta_T$ and $\theta_T + 0.1$ was equal to zero, and the proportion of early stoppings for safety was also equal to zero. These results are displayed in Table \ref{table_safety_operating_characteristics_spline}. In terms of the distribution of the recommended (true) utilities, Table \ref{table_utility_operating_characteristics_spline} shows that the average (true) utilities were close to the (true) utilities of the target dose combinations in all scenarios, showing that our design is able to capture complex uni-modal and bi-modal dose-utility surfaces. These average values were always above those obtained with the AAA design, with differences between 0.06 and 0.01. We observe that the median (true) utilities were also very close to the (true) utility of the target dose combination in all scenarios with practically no differences between the proposed design and the AAA design. One of the most notable differences between the proposed design and the AAA design was found in the 95\% confidence intervals of the distribution of recommended (true) utilities. The proposed design yielded, in general, narrower distributions, with notable differences with respect to the AAA design in scenarios 2, 4 and 6. We also see that the average (true) utility of the patients allocated with adaptive randomization was relatively close to the (true) utility of the target dose combination, taking into consideration that the adaptive randomization phase is still a learning part of the design. 

In Figure S1 of the supplementary material we display the (empirical) distribution of the recommended optimal dose combinations. We see how the design tends to recommend optimal dose combinations in the region(s) in which the (true) utility is higher. In scenarios 3, we observe a larger dispersion of the recommended optimal dose combinations with respect to the other scenarios, even though the average (true) utility is fairly close to the (true) utility of the target dose combinations. This dispersion is however expected given that this scenarios was designed to have a low number of positive efficacy outcomes.

\begin{table}
\caption{Safety operating characteristics from scenarios 1-6. The table displays the average DLT rate, the percentage of trials with DLT rates above $\theta_T$ and $\theta_T+0.1$, the percentage of trials stopped following the stage I and stage II safety stopping rules from each scenario.}
\centering
\resizebox{\columnwidth}{!}{%
\begin{tabular}{|c|cccccc|}
\hline
\multicolumn{1}{|c|}{\multirow{2}{*}{}} & \multicolumn{6}{c|}{Scenario} \\ \cline{2-7} 
\multicolumn{1}{|c|}{} & \multicolumn{1}{c|}{1} & \multicolumn{1}{c|}{2} & \multicolumn{1}{c|}{3} & \multicolumn{1}{c|}{4} & \multicolumn{1}{c|}{5} & 6 \\ \hline
\makecell{Average DLT rate (\%)} & \multicolumn{1}{c|}{12}  & \multicolumn{1}{c|}{17}  & \multicolumn{1}{c|}{7}  & \multicolumn{1}{c|}{0}  & \multicolumn{1}{c|}{11}  &  10 \\ \hline
\makecell{Percentage of trials with DLT rate above $\theta_T$ (\%)} & \multicolumn{1}{c|}{0}  & \multicolumn{1}{c|}{0}  & \multicolumn{1}{c|}{0}  & \multicolumn{1}{c|}{0}  & \multicolumn{1}{c|}{0}  &  0 \\ \hline
\makecell{Percentage of trials with DLT rate above $\theta_T + 0.1$ (\%)} & \multicolumn{1}{c|}{0}  & \multicolumn{1}{c|}{0}  & \multicolumn{1}{c|}{0}  & \multicolumn{1}{c|}{0}  & \multicolumn{1}{c|}{0}  & 0  \\ \hline
\makecell{Percentage of trials stopped early for safety (\%)} & \multicolumn{1}{c|}{0}  & \multicolumn{1}{c|}{0}  & \multicolumn{1}{c|}{0}  & \multicolumn{1}{c|}{0}  & \multicolumn{1}{c|}{0}  &  0 \\ \hline
\end{tabular}
}%
\label{table_safety_operating_characteristics_spline}
\end{table}

\begin{table}
\caption{Summary of the distribution of the (true) utility of the recommended optimal dose combinations from scenarios 1-6. The table displays the (true) utility of the target dose combination (TDC) as a reference, the average, median and 95\% confidence interval (i.e., percentiles 2.5\% and 97.5\%) of the distribution of (true) utility of the optimal dose combinations (ODC) recommended by the proposed design, the average, median and 95\% confidence interval (i.e., percentiles 2.5 and 97.5) of the distribution of (true) utility of the optimal dose combinations (ODC) recommended by the AAA design, and the average (true) utility of the patients allocated during the adaptive randomization (AR) phase from each scenario.}
\centering
\resizebox{\columnwidth}{!}{%
\begin{tabular}{|cc|cccccc|}
\hline
\multicolumn{2}{|c|}{\multirow{2}{*}{}} & \multicolumn{6}{c|}{Scenario} \\ \cline{3-8} 
\multicolumn{2}{|c|}{} & \multicolumn{1}{c|}{1} & \multicolumn{1}{c|}{2} & \multicolumn{1}{c|}{3} & \multicolumn{1}{c|}{4} & \multicolumn{1}{c|}{5} & 6 \\ \hline
\multicolumn{2}{|c|}{\makecell{Utility of TDC \\ (for reference)}} & \multicolumn{1}{c|}{\textbf{0.68}} & \multicolumn{1}{c|}{\textbf{0.94}} & \multicolumn{1}{c|}{\textbf{0.72}} & \multicolumn{1}{c|}{\textbf{0.98}} & \multicolumn{1}{c|}{\textbf{0.36}} & \textbf{0.39} \\ \hline
\multicolumn{1}{|c|}{\multirow{3}{*}{\makecell{Proposed \\ design}}}     & Average        & \multicolumn{1}{c|}{0.65}          & \multicolumn{1}{c|}{0.92}          & \multicolumn{1}{c|}{0.65}          & \multicolumn{1}{c|}{0.93}          & \multicolumn{1}{c|}{0.30}          & 0.33          \\ \cline{2-8} 
\multicolumn{1}{|c|}{}                                     & Median      & \multicolumn{1}{c|}{0.66}          & \multicolumn{1}{c|}{0.93}          & \multicolumn{1}{c|}{0.72}          & \multicolumn{1}{c|}{0.94}          & \multicolumn{1}{c|}{0.32}          & 0.34          \\ \cline{2-8} 
\multicolumn{1}{|c|}{}                                     & 95\% CI     & \multicolumn{1}{c|}{0.58-0.68}     & \multicolumn{1}{c|}{0.81-0.94}     & \multicolumn{1}{c|}{0.06-0.72}     & \multicolumn{1}{c|}{0.65-0.98}     & \multicolumn{1}{c|}{0.19-0.35}     & 0.21-0.38     \\ \hline
\multicolumn{1}{|c|}{\multirow{3}{*}{\makecell{AAA \\ design}}}          & Average        & \multicolumn{1}{c|}{0.62}          & \multicolumn{1}{c|}{0.86}          & \multicolumn{1}{c|}{0.60}          & \multicolumn{1}{c|}{0.89}          & \multicolumn{1}{c|}{0.28}          & 0.32          \\ \cline{2-8} 
\multicolumn{1}{|c|}{}                                     & Median      & \multicolumn{1}{c|}{0.66}          & \multicolumn{1}{c|}{0.93}          & \multicolumn{1}{c|}{0.72}          & \multicolumn{1}{c|}{0.94}          & \multicolumn{1}{c|}{0.30}          & 0.33          \\ \cline{2-8} 
\multicolumn{1}{|c|}{}                                     & 95\% CI     & \multicolumn{1}{c|}{0.55-0.66}     & \multicolumn{1}{c|}{0.40-0.94}     & \multicolumn{1}{c|}{0.06-0.72}     & \multicolumn{1}{c|}{0.57-0.98}     & \multicolumn{1}{c|}{0.16-0.33}     & 0.16-0.35     \\ \hline
\multicolumn{2}{|c|}{\makecell{Average utility of patients \\ allocated during AR phase}} & \multicolumn{1}{c|}{0.51}  & \multicolumn{1}{c|}{0.52}  & \multicolumn{1}{c|}{0.33}  & \multicolumn{1}{c|}{0.78}  & \multicolumn{1}{c|}{0.24}  &  0.26 \\ \hline
\end{tabular}
}%
\label{table_utility_operating_characteristics_spline}
\end{table}

\subsection{Robustness to deviations between the true underlying marginal models and the working marginal models}
\label{sc_robustness_model_deviations}

In this section we evaluated the performance of the proposed design by changing the true underlying marginal probability models. In other words, the binary toxicity and efficacy data were no longer generated using equations \eqref{eq_pdlt} and \eqref{eq_peff}. For this evaluation, we used the same marginal probability of toxicity and efficacy models from \cite{lyu2019aaa}. We note that these marginal probabilities do not include an interaction term modeling synergism between the two drugs following the recommendations of \cite{wang2005two} and the simulation results of \cite{Mozgunov2021} for discrete dose combinations. However, omitting an interaction term will compromise safety of the trial and reduce precision of the estimated MTD contour for continuous dose levels as shown in \cite{tighiouart2022}. We implemented ten out the eleven scenarios evaluated in their publication. We decided not to include scenario 4 (see Figure 5 in \cite{lyu2019aaa}) because the utility function was not defined for any of the dose combinations contained in the initial space of dose combination and, in this article, we restrict the dose-finding search to the initial space of standardized dose combinations $[0,1] \times [0,1]$. In a second evaluation, we implemented the proposed design in the six scenarios displayed in Figure \ref{plot_simulation_scenarios} with lower sample sizes in stages I and II. 

\begin{figure}
\caption{Contour plots showing the (true) utility surface from scenarios 1-3 and 5-11 from \cite{lyu2019aaa}. We also display the grid of discrete dose combination levels (i.e., black dots) with their corresponding (true) utility values whenever these (i.e., the utility values) are above zero.}
\centering
\vspace{0.25cm}
\includegraphics[scale=0.7]{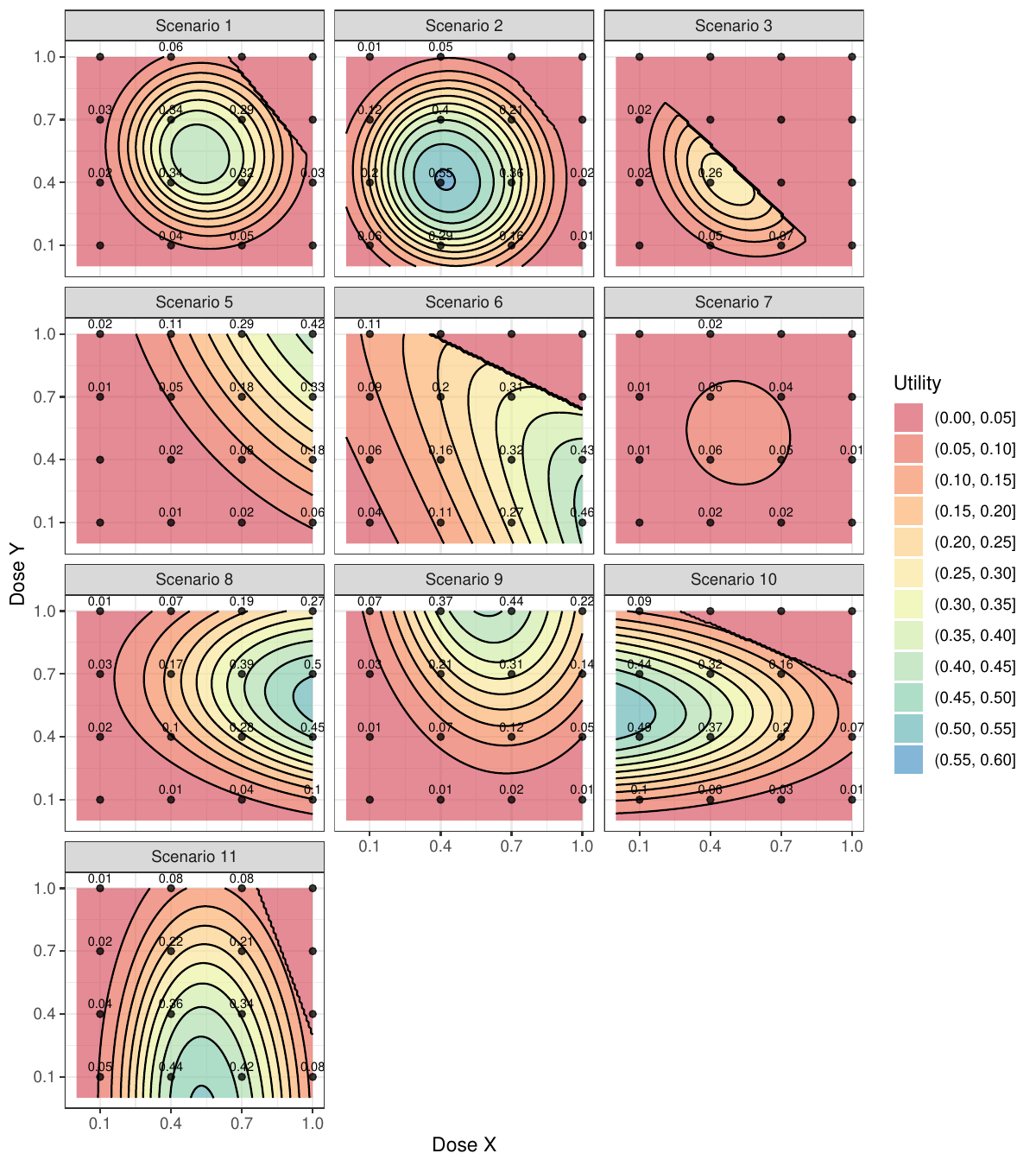}
\label{plot_simulation_scenarios_lyu}
\end{figure}

Regarding the first assessment (i.e., deviation between the true underlying marginal models and the working marginal models), in 
Figure \ref{plot_simulation_scenarios_lyu} we show the (true) utility surfaces for the simulated scenarios. The parameter values used to produce these scenarios are available in Table S2 of the supplementary material.

In terms of safety, the proposed design resulted in an average DLT rate between 13\% and 24\%, see Table \ref{table_safety_operating_characteristics_lyu}.  The percentage of trials with DLT rates above $\theta$ ranges between 0.0\% and 3.66\%, and the percentage of trial with DLT rates above $\theta_T + 0.1$ were equal to zero. Last, the percentage of trials stopped early for safety ranges between 0\% and 2.1\%. The safety results published by \cite{lyu2019aaa} show a worse performance with respect to our proposed design. The most notable difference is visible in scenario 3, where the MTD is located in the middle of the space of dose combinations. In this scenario, AAA design reported a proportion of trials with DLT rate above $\theta_T$ was 16.4\% whereas in our design this proportion is equal to 3.66\%. We would like to highlight that, from our point of view, a proportion of trials with DLT rate above $\theta_T$ of 16.4\% is too high and more restrictive safety criteria should be applied to lower it. 

In terms of the distribution of (true) recommended utilities, the proposed design has an average (true) utility that is, in general, close to the (true) utility of the target dose combinations, with values that are either very similar or higher than those of the AAA design under scenarios 1, 8, 10, and 11. The only scenario in which the proposed design seems to slightly underperform the AAA design is scenario 5. A similar behavior is observed with the medians of the recommended (true) utilities. The 95\% confidence interval of the distributions of the (true) utilities are, in general, narrower with the proposed design, with some notable differences under scenarios 1, 2, 8, 10, and 11. These results are consistent with those observed in the main simulation study of this manuscript (see Table \ref{table_utility_operating_characteristics_spline}). We also note that, in the proposed design, the average (true) utility of the patients allocated with adaptive randomization is relatively close to the (true) utility of the target dose combination, taking into consideration that the adaptive randomization phase is still a learning part of the design. These results are displayed in Table \ref{table_utility_operating_characteristics_lyu}.

In Figure S2 of the supplementary material we display the distribution of the recommended optimal dose combination in each scenario. Again, this Figure shows how the proposed design correctly identifies the region in which the target dose combination is located. Again, we notice that there is a higher dispersion in the optimal dose combination recommendations in scenarios in which the (true) utility of the target dose combination is, overall, not very high.

\begin{table}
\caption{Safety operating characteristics from scenarios 1-3 and 5-11 from \cite{lyu2019aaa}. The table displays the average DLT rate, the percentage of trials with DLT rates above $\theta_T$ of the proposed design, the percentage of trials with DLT rates above $\theta_T$ of the AAA design, and $\theta_T+0.1$, and the percentage of trials stopped following the stage I and stage II safety stopping rules from each scenario.}
\centering
\resizebox{\columnwidth}{!}{%
\begin{tabular}{|c|cccccccccc|}
\hline
\multicolumn{1}{|c|}{\multirow{2}{*}{}} & \multicolumn{10}{c|}{Scenario} \\ \cline{2-11} 
\multicolumn{1}{|c|}{} & \multicolumn{1}{c|}{1} & \multicolumn{1}{c|}{2} & \multicolumn{1}{c|}{3} & \multicolumn{1}{c|}{5} & \multicolumn{1}{c|}{6} & \multicolumn{1}{c|}{7} & \multicolumn{1}{c|}{8} & \multicolumn{1}{c|}{9} & \multicolumn{1}{c|}{10} &  11 \\ \hline
\makecell{Average DLT rate (\%) \\ (proposed)} & \multicolumn{1}{c|}{17}  & \multicolumn{1}{c|}{16}  & \multicolumn{1}{c|}{24}  & \multicolumn{1}{c|}{18}  & \multicolumn{1}{c|}{19} & \multicolumn{1}{c|}{17}  & \multicolumn{1}{c|}{13}  & \multicolumn{1}{c|}{13}  & \multicolumn{1}{c|}{17}  &  18 \\ \hline
\makecell{Percentage of trials with \\ DLT rate above $\theta_T$ (\%) (proposed)} & \multicolumn{1}{c|}{0}  & \multicolumn{1}{c|}{0}  & \multicolumn{1}{c|}{\bf{3.66}}  & \multicolumn{1}{c|}{0.05}  &  \multicolumn{1}{c|}{0.10} & \multicolumn{1}{c|}{0}  & \multicolumn{1}{c|}{0}  & \multicolumn{1}{c|}{0}  & \multicolumn{1}{c|}{0}  &  0 \\ \hline
\makecell{Percentage of trials with \\ DLT rate above $\theta_T$ (\%) (AAA)} & \multicolumn{1}{c|}{0}  & \multicolumn{1}{c|}{0.1}  & \multicolumn{1}{c|}{\bf{16.4}}  & \multicolumn{1}{c|}{0.3}  &  \multicolumn{1}{c|}{1.3} & \multicolumn{1}{c|}{0}  & \multicolumn{1}{c|}{0}  & \multicolumn{1}{c|}{0}  & \multicolumn{1}{c|}{0}  & 0.1  \\ \hline
\makecell{Percentage of trials with \\ DLT rate above $\theta_T + 0.1$ (\%) (proposed)} & \multicolumn{1}{c|}{0}  & \multicolumn{1}{c|}{0}  & \multicolumn{1}{c|}{0}  & \multicolumn{1}{c|}{0}  & \multicolumn{1}{c|}{0}  & \multicolumn{1}{c|}{0}  & \multicolumn{1}{c|}{0} & \multicolumn{1}{c|}{0}  & \multicolumn{1}{c|}{0}  &  0  \\ \hline
\makecell{Percentage of trials \\ stopped early for safety (\%) (proposed)} & \multicolumn{1}{c|}{0.25} & \multicolumn{1}{c|}{0.15}  & \multicolumn{1}{c|}{0.40}  & \multicolumn{1}{c|}{2.10}  & \multicolumn{1}{c|}{0.40}  & \multicolumn{1}{c|}{0.35}  & \multicolumn{1}{c|}{0.10}  & \multicolumn{1}{c|}{0.05}  & \multicolumn{1}{c|}{0.40}  & 0.85 \\ \hline
\end{tabular}
}
\label{table_safety_operating_characteristics_lyu}
\end{table}

\begin{table}
\caption{Summary of the distribution of the (true) utility of the recommended optimal dose combinations from scenarios 1-3 and 5-11 from \cite{lyu2019aaa}. The table displays the (true) utility of the target dose combination (TDC) as a reference, the average, median and 95\% confidence interval (i.e., percentiles 2.5\% and 97.5\%) of the distribution of (true) utility of the optimal dose combinations (ODC) recommended by the proposed design, the average, median and 95\% confidence interval (i.e., percentiles 2.5 and 97.5) of the distribution of (true) utility of the optimal dose combinations (ODC) recommended by the AAA design, and the average (true) utility of the patients allocated during the adaptive randomization (AR) phase from each scenario.}
\resizebox{\columnwidth}{!}{%
\begin{tabular}{|cc|cccccccccc|}
\hline
\multicolumn{2}{|c|}{\multirow{2}{*}{}} & \multicolumn{10}{c|}{Scenario} \\ \cline{3-12} 
\multicolumn{2}{|c|}{} & \multicolumn{1}{c|}{1} & \multicolumn{1}{c|}{2} & \multicolumn{1}{c|}{3} & \multicolumn{1}{c|}{5} & \multicolumn{1}{c|}{6} & \multicolumn{1}{c|}{7} & \multicolumn{1}{c|}{8} & \multicolumn{1}{c|}{9} & \multicolumn{1}{c|}{10} & 11 \\ \hline
\multicolumn{2}{|c|}{\makecell{Utility of TDC \\ (for reference)}} & \multicolumn{1}{c|}{\textbf{0.45}} & \multicolumn{1}{c|}{\textbf{0.55}} & \multicolumn{1}{c|}{\textbf{0.28}} & \multicolumn{1}{c|}{\textbf{0.42}} & \multicolumn{1}{c|}{\textbf{0.47}} & \multicolumn{1}{c|}{\textbf{0.07}} & \multicolumn{1}{c|}{\textbf{0.52}} & \multicolumn{1}{c|}{\textbf{0.46}} & \multicolumn{1}{c|}{\textbf{0.55}} & \textbf{0.51} \\ \hline
\multicolumn{1}{|c|}{\multirow{3}{*}{\makecell{Proposed \\ design}}} & Average & \multicolumn{1}{c|}{0.42} & \multicolumn{1}{c|}{0.53} & \multicolumn{1}{c|}{0.20} & \multicolumn{1}{c|}{0.26} & \multicolumn{1}{c|}{0.38} & \multicolumn{1}{c|}{0.05} & \multicolumn{1}{c|}{0.46} & \multicolumn{1}{c|}{0.39} & \multicolumn{1}{c|}{0.52} & 0.48 \\ \cline{2-12} 
\multicolumn{1}{|c|}{} & Median & \multicolumn{1}{c|}{0.43} & \multicolumn{1}{c|}{0.54} & \multicolumn{1}{c|}{0.26} & \multicolumn{1}{c|}{0.29} & \multicolumn{1}{c|}{0.45} & \multicolumn{1}{c|}{0.06} & \multicolumn{1}{c|}{0.48} & \multicolumn{1}{c|}{0.42} & \multicolumn{1}{c|}{0.54} & 0.50 \\ \cline{2-12} 
\multicolumn{1}{|c|}{} & 95\% CI & \multicolumn{1}{c|}{0.35-0.45} & \multicolumn{1}{c|}{0.45-0.55} & \multicolumn{1}{c|}{0.00-0.28} & \multicolumn{1}{c|}{0.01-0.42} & \multicolumn{1}{c|}{0.00-0.47} & \multicolumn{1}{c|}{0.00-0.07} & \multicolumn{1}{c|}{0.21-0.52} & \multicolumn{1}{c|}{0.13-0.46} & \multicolumn{1}{c|}{0.41-0.55} & 0.40-0.51 \\ \hline
\multicolumn{1}{|c|}{\multirow{3}{*}{\makecell{AAA \\ design}}} & Average & \multicolumn{1}{c|}{0.37} & \multicolumn{1}{c|}{0.53} & \multicolumn{1}{c|}{0.20} & \multicolumn{1}{c|}{0.31} & \multicolumn{1}{c|}{0.37} & \multicolumn{1}{c|}{0.04} & \multicolumn{1}{c|}{0.44} & \multicolumn{1}{c|}{0.39} & \multicolumn{1}{c|}{0.46} & 0.43 \\ \cline{2-12} 
\multicolumn{1}{|c|}{} & Median & \multicolumn{1}{c|}{0.34} & \multicolumn{1}{c|}{0.55} & \multicolumn{1}{c|}{0.26} & \multicolumn{1}{c|}{0.33} & \multicolumn{1}{c|}{0.46} & \multicolumn{1}{c|}{0.05} & \multicolumn{1}{c|}{0.45} & \multicolumn{1}{c|}{0.44} & \multicolumn{1}{c|}{0.49} & 0.44 \\ \cline{2-12} 
\multicolumn{1}{|c|}{} & 95\% CI & \multicolumn{1}{c|}{0.29-0.45} & \multicolumn{1}{c|}{0.29-0.55} & \multicolumn{1}{c|}{0.00-0.28} & \multicolumn{1}{c|}{0.01-0.42} & \multicolumn{1}{c|}{0.00-0.46} & \multicolumn{1}{c|}{0.00-0.07} & \multicolumn{1}{c|}{0.17-0.52} & \multicolumn{1}{c|}{0.14-0.46} & \multicolumn{1}{c|}{0.32-0.53} & 0.34-0.50 \\ \hline
\multicolumn{2}{|c|}{\makecell{Average utility \\ of patients allocated \\ during AR phase}} & \multicolumn{1}{c|}{0.24} & \multicolumn{1}{c|}{0.31} & \multicolumn{1}{c|}{0.12} & \multicolumn{1}{c|}{0.16} & \multicolumn{1}{c|}{0.26} & \multicolumn{1}{c|}{0.03} & \multicolumn{1}{c|}{0.29} & \multicolumn{1}{c|}{0.24} & \multicolumn{1}{c|}{0.31} & 0.31 \\ \hline
\end{tabular}%
}
\label{table_utility_operating_characteristics_lyu}
\end{table}

\subsection{Robustness to changes in the sample size}
In this section, we re-assessed the proposed design in the six scenarios generated with the proposed working marginal probability models (i.e., Figure \ref{plot_simulation_scenarios}) with lower sample sizes in stages I and II. The results are available in Tables S3 and S4 of the supplementary material for sample size of 10, 20, and 30 for stage I, 30, 42, 54, and 66 in stage II, and cohort size 6 and 12 in stage II. Overall, we observed that reducing the number of patients only in stage II  yields worse performance than reducing patients only in stage I. We also observed that, for all sample sizes considered in stage I, all scenarios with 54 and 42 patients in stage II had a performance that was still as good as or better than the performance obtained with the AAA design, and the only case in which the proposed design was slightly worse, in some scenarios, than the AAA design, was when stage II uses only 30 patients. In other words, we could enroll 62 patients (i.e., 20 in stage I and 42 in stage II), which would be a 35.42\% reduction of the overall sample size, and still have results as good as or better than those from the AAA design.


\subsection{\com{Robustness to independence between the toxicity and efficacy marginal probabilities}}

\com{In this section, we assess the performance of the design when we assume a correlation structure between the toxicity and efficacy marginals. To do so, we sample latent variables $(L_{T},L_{E})$ from a bivariate normal distribution with means $\mu_T = \alpha_0 + \alpha_1 x + \alpha_2 y + \eta x y$ and $\mu_E = \beta_0 + \beta_1 x + \beta_2 x^2 + \sum_{i=3}^{5} \beta_i (x - \kappa_{i-2})_{+}^{3} + \beta_6 y + \beta_7 y^2 + \sum_{j=8}^{10} \beta_j (y - \kappa_{j-4})_{+}^{3} + \beta_{11}xy$, standard deviations $\sigma_T = \sigma_E = 1$, and different correlation coefficients $\lambda = 0, 0.25, 0.5, 0.75, 1$. The DLT and efficacy outcomes are defined as $Z_{T} = I(L_{T} > 0)$ and $Z_{E} = I(L_{E} > 0)$, respectively. The reason why we include $\lambda = 0$ (i.e., no correlation) is because the data generation model in this assessment is different from the one used throughout the article.}

\com{We performed simulation studies under Scenarios 1 and 3 in \cite{lyu2019aaa}. In general, the differences in the operating characteristics are negligible with respect to both safety and the recommended dose combinations across the different correlation coefficients. The results are presented in Table S5 in the supplementary material. These results are consistent with the findings in \cite{cai2014bayesian,cunanan2014} where a similar assessment was made with the Gumbel copula model. However, we note that this assessment is limited by using a bivariate latent variable model to induce a correlation structure. Other joint distributions (e.g. copula models) could be used for further exploration, although this is beyond the scope of this article}.

\section{Discussion}
\label{sc_discussion}

We proposed a Bayesian phase I-II design for dual-agent dose optimization with molecularly targeted therapies. In the presence of these types of compounds, the monotonicity assumption of the dose-efficacy response may not be appropriate given that efficacy may decrease or plateau at high dose levels, and thus the optimal dose will be a trade-off between toxicity and efficacy. For this reason, we employ a flexible cubic spline function for the marginal distribution of the efficacy response and a utility function to assess the risk-benefit trade-off between undesirable and desirable clinical outcomes. The proposed marginal dose-efficacy model has a relatively high number of parameters, which accommodates simple and complex (e.g., bi-modal) dose-efficacy surfaces. However, because in stage I dose-escalation is driven by the marginal dose-toxicity model, which has a small number of parameters, and we do not allow for skipping untried dose combinations, there is no concern regarding the variability of the parameter estimation at the inception of the trial. Then, by the time the design starts using the marginal probability of efficacy to allocate subsequent cohorts of patients, there are already enough patients enrolled in the trial. 

Traditionally, in settings in which the assumption of monotonicity of the dose-efficacy holds (i.e., with cytotoxic agents), the purpose was not to have a precise estimation of the entire dose-efficacy curve (or surface) but to provide a precise estimation of the MTD in phase I trials, or the optimal dose in phase I-II trials. However, without the assumption of monotonicity of the dose-efficacy curve (or surface), implementing dose optimization becomes harder since the target dose combination could be anywhere in the space of doses available in the clinical trial and in regions with low probability of toxicity. In this situation, having a more precise estimation of the entire dose-efficacy curve (or surface) could be helpful since we can better locate the region(s) in which the target dose (or dose combination) is located.

We are inspired by a published case study which combines a MEK inhibitor and a PIK3CA inhibitor considering 4 discrete dose levels in each compound and a total of 96 late-stage cancer patients (\cite{lyu2019aaa}). In this setting, we first implemented our approach using 6 scenarios generated by the working marginal dose-toxicity and dose-efficacy models. These scenarios were intended to have different utility contour shapes, with one or two modes, and with low, medium and high (true) utility values for the target dose combinations. The proposed design achieved good operating characteristics, with low toxicity rates and with average recommended (true) utilities that were close to the (true) utilities of the target dose combinations. In other words, the design was, on average, able to correctly identify the region(s) of the utility surface in which the target dose combination was located.

To assess model robustness to deviations from the true marginal models for toxicity and efficacy, we derived operating characteristics under the scenarios presented in \cite{lyu2019aaa}. The proposed design showed, again, low toxicity rates and average (true) utilities of the recommended optimal dose combination that were close to the (true) utilities of the scenario-specific target dose combinations.

The performance of the proposed design was compared to that of the AAA design, a state-of-the-art design proposed by \cite{lyu2019aaa} specifically tailored for the combination of molecularly targeted therapies. For this purpose, we implemented the AAA design in the scenarios generated with the working marginal models that we proposed in this article, as well as in the scenarios generated with the marginal models from \cite{lyu2019aaa}. In terms of safety, we have observed that the proposed design is safer than the AAA design, as explained in section \ref{sc_robustness_model_deviations}. In terms of target dose combination(s) recommendation, we looked at the distribution of the (true) utilities of the recommended optimal dose combinations produced by both designs. These results showed that, under the presence of more complex (e.g., bi-modal) utility surfaces, the proposed design had better performance (i.e., higher average (true) utility of the recommended optimal dose combinations) than the AAA design. When the surfaces were less complex (e.g., uni-modal), the differences in performance between the proposed design and the AAA design became smaller. The distribution of the (true) utilities of the recommended optimal dose combinations of the proposed design, measured through the percentiles 2.5 and 97.5, were narrower with respect to those from the AAA design, with notable differences in some scenarios. We also evaluated the robustness of the proposed design with respect to smaller sample sizes in stages I and II. This assessment showed that the operating characteristics of the design with smaller sample sizes in both stages were still close to those obtained with the sample size used for the main simulation study. As a reference, we observed that an overall sample size reduction of 35\% still leads to operating characteristics that were as good as or better than those produced by the AAA design in the evaluated scenarios. With these results, we conclude that the number of parameters employed by the marginal probability of efficacy model is not an issue in this type of clinical trial setting.

In general, we believe that the proposed design has the following advantages over the AAA design: i) it has better safety operating characteristics, ii) it has better performance, specially under complex dose-utility surfaces, iii) it is more precise in terms of optimal dose combination recommendation, and iv) it is easier to implement since it does not incorporate the adaptive dose insertion feature of the AAA design, which could be challenging in practice if many dose insertions are needed during the trial. One potential weakness of the proposed design has been observed in the scenarios in which the (true) utility surface had, overall, low values. In this setting, there was a slightly higher dispersion of the optimal dose combination recommendation with respect to other tested scenarios, which translated into slightly lower average (true) utilities. However, a higher dispersion in this situation was not completely unexpected since low utility values are usually a result of low efficacy rates, which, given the relatively high number of parameter of the marginal efficacy model, can lead to lower estimation precision. Nevertheless, because we are using vague prior distributions, we can consider the results presented in this manuscript as the baseline operating characteristics of the proposed design. By using more informative prior distributions in the marginal probability models, and/or increasing the sample size, the design’s operating characteristics are expected to improve.

\subsection*{Data availability}
No real data was used in the development of this article. The \texttt{R} and \texttt{JAGS} scripts necessary to fully reproduce the results presented in this article are available at \newline
\url{https://github.com/jjimenezm1989/Phase-I-II-design-targeted-therapies}.

\subsection*{Funding}

Mourad Tighiouart is funded by NIH the National Center for Advancing Translational Sciences (NCATS) UCLA CTSI (UL1 TR001881-01), NCI P01 CA233452-02, and U01 grant CA232859-01.

\subsection*{Disclaimer}

Jos\'e L. Jim\'enez is employed by Novartis Pharma A.G. who provided support in the form of salary for the author, but did not have any additional role in the preparation of the manuscript. Also, the views expressed in this publication are those of the authors and should not be attributed to any of the funding institutions or organisations to which the authors are affiliated.

\bibliographystyle{plainnat} 
\bibliography{references.bib}

\begin{thebibliography}{43}
\providecommand{\natexlab}[1]{#1}
\providecommand{\url}[1]{\texttt{#1}}
\expandafter\ifx\csname urlstyle\endcsname\relax
  \providecommand{\doi}[1]{doi: #1}\else
  \providecommand{\doi}{doi: \begingroup \urlstyle{rm}\Url}\fi

\bibitem[Babb et~al.(1998)Babb, Rogatko, and Zacks]{babb1998cancer}
James Babb, Andr{\'e} Rogatko, and Shelemyahu Zacks.
\newblock Cancer phase {I} clinical trials: efficient dose escalation with
  overdose control.
\newblock \emph{Statistics in Medicine}, 17\penalty0 (10):\penalty0 1103--1120,
  1998.

\bibitem[Cai et~al.(2014)Cai, Yuan, and Ji]{cai2014bayesian}
Chunyan Cai, Ying Yuan, and Yuan Ji.
\newblock A {Bayesian} dose-finding design for oncology clinical trials of
  combinational biological agents.
\newblock \emph{Journal of the Royal Statistical Society. Series C, Applied
  statistics}, 63\penalty0 (1):\penalty0 159, 2014.

\bibitem[Cunanan and Koopmeiners(2014)]{cunanan2014}
Kristen Cunanan and Joseph~S Koopmeiners.
\newblock Evaluating the performance of copula models in phase i-ii clinical
  trials under model misspecification.
\newblock \emph{BMC Medical Research Methodology}, 14:\penalty0 51, 2014.

\bibitem[Diniz et~al.(2017)Diniz, Li, and Tighiouart]{diniz2017}
M{\'a}rcio~Augusto Diniz, Quanlin Li, and Mourad Tighiouart.
\newblock Dose finding for drug combination in early cancer phase {I} trials
  using conditional continual reassessment method.
\newblock \emph{Journal of Biometrics \& Biostatistics}, 8\penalty0 (6), 2017.

\bibitem[Diniz et~al.(2018)Diniz, Kim, and Tighiouart]{diniz2018}
M{\'a}rcio~Augusto Diniz, Sungjin Kim, and Mourad Tighiouart.
\newblock A {Bayesian} adaptive design in cancer phase {I} trials using dose
  combinations in the presence of a baseline covariate.
\newblock \emph{Journal of Probability and Statistics}, 2018, 2018.

\bibitem[Diniz et~al.(2019)Diniz, Tighiouart, and Rogatko]{diniz2019comparison}
M{\'a}rcio~Augusto Diniz, Mourad Tighiouart, and Andr{\'e} Rogatko.
\newblock Comparison between continuous and discrete doses for model based
  designs in cancer dose finding.
\newblock \emph{PLoS One}, 14\penalty0 (1):\penalty0 e0210139, 2019.

\bibitem[Guo and Yuan(2015)]{guo2015bayesian}
Beibei Guo and Ying Yuan.
\newblock A {Bayesian} dose-finding design for phase {I/II} clinical trials
  with nonignorable dropouts.
\newblock \emph{Statistics in Medicine}, 34\penalty0 (10):\penalty0 1721--1732,
  2015.

\bibitem[Guo and Yuan(2017)]{guo2017bayesian}
Beibei Guo and Ying Yuan.
\newblock Bayesian phase {I/II} biomarker-based dose finding for precision
  medicine with molecularly targeted agents.
\newblock \emph{Journal of the American Statistical Association}, 112\penalty0
  (518):\penalty0 508--520, 2017.

\bibitem[Hoff and Ellis(2007)]{hoff2007targeted}
Paulo~M Hoff and Lee~M Ellis.
\newblock Targeted therapy trials: Approval strategies, target validation, or
  helping patients?
\newblock \emph{Journal of Clinical Oncology}, 25\penalty0 (13):\penalty0
  1639--1641, 2007.

\bibitem[Houede et~al.(2010)Houede, Thall, Nguyen, Paoletti, and
  Kramar]{houede2010utility}
Nadine Houede, Peter~F Thall, Hoang Nguyen, Xavier Paoletti, and Andrew Kramar.
\newblock Utility-based optimization of combination therapy using ordinal
  toxicity and efficacy in phase {I/II} trials.
\newblock \emph{Biometrics}, 66\penalty0 (2):\penalty0 532--540, 2010.

\bibitem[Ivanova et~al.(2009)Ivanova, Liu, Snyder, and
  Snavely]{ivanova2009adaptive}
Anastasia Ivanova, Ken Liu, Ellen Snyder, and Duane Snavely.
\newblock An adaptive design for identifying the dose with the best
  efficacy/tolerability profile with application to a crossover dose-finding
  study.
\newblock \emph{Statistics in Medicine}, 28\penalty0 (24):\penalty0 2941--2951,
  2009.

\bibitem[Jim{\'e}nez and Tighiouart(2022)]{jimenez2021combining}
Jos{\'e}~L Jim{\'e}nez and Mourad Tighiouart.
\newblock Combining cytotoxic agents with continuous dose levels in seamless
  phase {I-II} clinical trials.
\newblock \emph{Journal of the Royal Statistical Society: Series C (Applied
  Statistics)}, 71\penalty0 (5):\penalty0 1996--2013, 2022.

\bibitem[Jim{\'e}nez and Zheng(2021)]{jimenez2021bayesian}
Jos{\'e}~L Jim{\'e}nez and Haiyan Zheng.
\newblock A {Bayesian} adaptive design for dual-agent phase {I-II} cancer
  clinical trials combining efficacy data across stages.
\newblock \emph{arXiv preprint}, 2021.

\bibitem[Jimenez et~al.(2019)Jimenez, Tighiouart, and Gasparini]{jimenez2019}
Jose~L Jimenez, Mourad Tighiouart, and Mauro Gasparini.
\newblock Cancer phase {I} trial design using drug combinations when a fraction
  of dose limiting toxicities is attributable to one or more agents.
\newblock \emph{Biometrical Journal}, 61\penalty0 (2):\penalty0 319--332, 2019.

\bibitem[Jim{\'e}nez et~al.(2020)Jim{\'e}nez, Kim, and
  Tighiouart]{jimenez2020bayesian}
Jos{\'e}~L Jim{\'e}nez, Sungjin Kim, and Mourad Tighiouart.
\newblock A bayesian seamless phase {I--II} trial design with two stages for
  cancer clinical trials with drug combinations.
\newblock \emph{Biometrical Journal}, 62\penalty0 (5):\penalty0 1300--1314,
  2020.

\bibitem[Le~Tourneau et~al.(2009)Le~Tourneau, Lee, and Siu]{le2009dose}
Christophe Le~Tourneau, J~Jack Lee, and Lillian~L Siu.
\newblock Dose escalation methods in phase {I} cancer clinical trials.
\newblock \emph{JNCI: Journal of the National Cancer Institute}, 101\penalty0
  (10):\penalty0 708--720, 2009.

\bibitem[Li et~al.(2017)Li, Whitmore, Guo, and Ji]{li2017toxicity}
Daniel~H Li, James~B Whitmore, Wentian Guo, and Yuan Ji.
\newblock Toxicity and efficacy probability interval design for phase {I}
  adoptive cell therapy dose-finding clinical trials.
\newblock \emph{Clinical Cancer Research}, 23\penalty0 (1):\penalty0 13--20,
  2017.

\bibitem[Lin et~al.(2020{\natexlab{a}})Lin, Thall, and Yuan]{lin2020adaptive}
Ruitao Lin, Peter~F Thall, and Ying Yuan.
\newblock An adaptive trial design to optimize dose-schedule regimes with
  delayed outcomes.
\newblock \emph{Biometrics}, 76\penalty0 (1):\penalty0 304--315,
  2020{\natexlab{a}}.

\bibitem[Lin et~al.(2020{\natexlab{b}})Lin, Zhou, Yan, Li, and
  Yuan]{lin2020boin12}
Ruitao Lin, Yanhong Zhou, Fangrong Yan, Daniel Li, and Ying Yuan.
\newblock Boin12: Bayesian optimal interval phase i/ii trial design for
  utility-based dose finding in immunotherapy and targeted therapies.
\newblock \emph{JCO precision oncology}, 4:\penalty0 1393--1402,
  2020{\natexlab{b}}.

\bibitem[Liu et~al.(2018)Liu, Guo, and Yuan]{liu2018bayesian}
Suyu Liu, Beibei Guo, and Ying Yuan.
\newblock A {Bayesian} phase {I/II} trial design for immunotherapy.
\newblock \emph{Journal of the American Statistical Association}, 113\penalty0
  (523):\penalty0 1016--1027, 2018.

\bibitem[Lu et~al.(2024)Lu, Zhang, Yuan, and Lin]{lu2024comb}
Mengyi Lu, Jingyi Zhang, Ying Yuan, and Ruitao Lin.
\newblock Comb-boin12: A utility-based bayesian optimal interval design for
  dose optimization in cancer drug-combination trials.
\newblock \emph{Statistics in Biopharmaceutical Research}, \penalty0
  (just-accepted):\penalty0 1--22, 2024.

\bibitem[Lyu et~al.(2019)Lyu, Ji, Zhao, and Catenacci]{lyu2019aaa}
Jiaying Lyu, Yuan Ji, Naiqing Zhao, and Daniel~VT Catenacci.
\newblock {AAA}: triple adaptive {Bayesian} designs for the identification of
  optimal dose combinations in dual-agent dose finding trials.
\newblock \emph{Journal of the Royal Statistical Society. Series C, Applied
  statistics}, 68\penalty0 (2):\penalty0 385, 2019.

\bibitem[Mozgunov et~al.(2021)Mozgunov, Knight, Barnett, and
  Jaki]{Mozgunov2021}
Pavel Mozgunov, R~Knight, H~Barnett, and Thomas Jaki.
\newblock Using an interaction parameter in model-based phase i trials for
  combination treatments? a simulation study.
\newblock \emph{Int J Environ Res Public Health}, 18\penalty0 (1):\penalty0
  345, 2021.

\bibitem[Murray et~al.(2017)Murray, Thall, Yuan, McAvoy, and
  Gomez]{murray2017robust}
Thomas~A Murray, Peter~F Thall, Ying Yuan, Sarah McAvoy, and Daniel~R Gomez.
\newblock Robust treatment comparison based on utilities of semi-competing
  risks in non-small-cell lung cancer.
\newblock \emph{Journal of the American Statistical Association}, 112\penalty0
  (517):\penalty0 11--23, 2017.

\bibitem[Nebiyou~Bekele and Shen(2005)]{nebiyou2005bayesian}
B~Nebiyou~Bekele and Yu~Shen.
\newblock A {Bayesian} approach to jointly modeling toxicity and biomarker
  expression in a phase {I/II} dose-finding trial.
\newblock \emph{Biometrics}, 61\penalty0 (2):\penalty0 343--354, 2005.

\bibitem[Plummer et~al.(2003)]{plummer2003jags}
Martyn Plummer et~al.
\newblock Jags: A program for analysis of {Bayesian} graphical models using
  gibbs sampling.
\newblock In \emph{Proceedings of the 3rd international workshop on distributed
  statistical computing}, volume 124, pages 1--10. Vienna, Austria., 2003.

\bibitem[{R Core Team}(2019)]{r2019}
{R Core Team}.
\newblock \emph{R: A Language and Environment for Statistical Computing}.
\newblock R Foundation for Statistical Computing, Vienna, Austria, 2019.
\newblock URL \url{https://www.R-project.org/}.

\bibitem[Shi and Yin(2013)]{shi2013escalation}
Yun Shi and Guosheng Yin.
\newblock Escalation with overdose control for phase {I} drug-combination
  trials.
\newblock \emph{Statistics in Medicine}, 32\penalty0 (25):\penalty0 4400--4412,
  2013.

\bibitem[Thall and Cook(2004)]{thall2004dose}
Peter~F Thall and John~D Cook.
\newblock Dose-finding based on efficacy-toxicity trade-offs.
\newblock \emph{Biometrics}, 60\penalty0 (3):\penalty0 684--693, 2004.

\bibitem[Thall et~al.(2013)Thall, Nguyen, Braun, and Qazilbash]{thall2013using}
Peter~F Thall, Hoang~Q Nguyen, Thomas~M Braun, and Muzaffar~H Qazilbash.
\newblock Using joint utilities of the times to response and toxicity to
  adaptively optimize schedule--dose regimes.
\newblock \emph{Biometrics}, 69\penalty0 (3):\penalty0 673--682, 2013.

\bibitem[Thall et~al.(2014)Thall, Nguyen, Zohar, and
  Maton]{thall2014optimizing}
Peter~F Thall, Hoang~Q Nguyen, Sarah Zohar, and Pierre Maton.
\newblock Optimizing sedative dose in preterm infants undergoing treatment for
  respiratory distress syndrome.
\newblock \emph{Journal of the American Statistical Association}, 109\penalty0
  (507):\penalty0 931--943, 2014.

\bibitem[Tighiouart(2019)]{tighiouart2019two}
Mourad Tighiouart.
\newblock Two-stage design for phase {I--II} cancer clinical trials using
  continuous dose combinations of cytotoxic agents.
\newblock \emph{Journal of the Royal Statistical Society: Series C (Applied
  Statistics)}, 68\penalty0 (1):\penalty0 235--250, 2019.

\bibitem[Tighiouart and Rogatko(2010)]{tighiouart2010dose}
Mourad Tighiouart and Andr{\'e} Rogatko.
\newblock Dose finding with escalation with overdose control ({EWOC}) in cancer
  clinical trials.
\newblock \emph{Statistical Science}, 25\penalty0 (2):\penalty0 217--226, 2010.

\bibitem[Tighiouart and Rogatko(2012)]{tighiouart2012number}
Mourad Tighiouart and Andre Rogatko.
\newblock Number of patients per cohort and sample size considerations using
  dose escalation with overdose control.
\newblock \emph{Journal of Probability and Statistics}, 2012, 2012.

\bibitem[Tighiouart et~al.(2005)Tighiouart, Rogatko, and
  Babb]{tighiouart2005flexible}
Mourad Tighiouart, Andr{\'e} Rogatko, and James~S Babb.
\newblock Flexible {Bayesian} methods for cancer phase {I} clinical trials.
  dose escalation with overdose control.
\newblock \emph{Statistics in Medicine}, 24\penalty0 (14):\penalty0 2183--2196,
  2005.

\bibitem[Tighiouart et~al.(2014)Tighiouart, Piantadosi, and
  Rogatko]{tighiouart2014}
Mourad Tighiouart, Steven Piantadosi, and Andr{\'{e}} Rogatko.
\newblock {Dose finding with drug combinations in cancer phase {I} clinical
  trials using conditional escalation with overdose control}.
\newblock \emph{Statistics in Medicine}, 33\penalty0 (22):\penalty0 3815--29,
  2014.

\bibitem[Tighiouart et~al.(2017)Tighiouart, Li, and
  Rogatko]{tighiouart2017bayesian}
Mourad Tighiouart, Quanlin Li, and Andr{\'e} Rogatko.
\newblock A {Bayesian} adaptive design for estimating the maximum tolerated
  dose curve using drug combinations in cancer phase {I} clinical trials.
\newblock \emph{Statistics in Medicine}, 36\penalty0 (2):\penalty0 280--290,
  2017.

\bibitem[Tighiouart et~al.(2022)Tighiouart, Jim\'enez, Diniz, and
  Rogatko]{tighiouart2022}
Mourad Tighiouart, Jos\'e~L Jim\'enez, Marcio~A Diniz, and Andr\'e Rogatko.
\newblock Tmodeling synergism in early phase cancer trials with drug
  combination with continuous dose levels: is there an added value?
\newblock \emph{Brazilian Journal of Biometrics, in print}, 2022.

\bibitem[Wang and Ivanova(2005)]{wang2005two}
Kai Wang and Anastasia Ivanova.
\newblock Two-dimensional dose finding in discrete dose space.
\newblock \emph{Biometrics}, 61\penalty0 (1):\penalty0 217--222, 2005.

\bibitem[Wheeler et~al.(2017)Wheeler, Sweeting, and
  Mander]{wheeler2017toxicity}
Graham~M Wheeler, Michael~J Sweeting, and Adrian~P Mander.
\newblock Toxicity-dependent feasibility bounds for the escalation with
  overdose control approach in phase {I} cancer trials.
\newblock \emph{Statistics in Medicine}, 36\penalty0 (16):\penalty0 2499--2513,
  2017.

\bibitem[Yuan and Yin(2011)]{yuan2011bayesian}
Ying Yuan and Guosheng Yin.
\newblock Bayesian phase {I/II} adaptively randomized oncology trials with
  combined drugs.
\newblock \emph{The Annals of Applied Statistics}, 5\penalty0 (2A):\penalty0
  924, 2011.

\bibitem[Yuan et~al.(2016)Yuan, Nguyen, and Thall]{yuan2016bayesian}
Ying Yuan, Hoang~Q Nguyen, and Peter~F Thall.
\newblock \emph{Bayesian designs for phase I--II clinical trials}.
\newblock CRC Press, 2016.

\bibitem[Zhang et~al.(2006)Zhang, Sargent, and Mandrekar]{zhang2006adaptive}
Wei Zhang, Daniel~J Sargent, and Sumithra Mandrekar.
\newblock An adaptive dose-finding design incorporating both toxicity and
  efficacy.
\newblock \emph{Statistics in Medicine}, 25\penalty0 (14):\penalty0 2365--2383,
  2006.

\end{thebibliography}

\end{document}